\documentclass[1p,onecolumn]{elsarticle}
\usepackage{amssymb}
\usepackage{graphicx}
\usepackage{dcolumn}
\usepackage{bm}
\usepackage[cdot,mediumqspace,amssymb]{SIunits} 
\usepackage[autolanguage]{numprint} 
\usepackage{ifthen}
\usepackage{ifpdf}
\usepackage{wrapfig}
\ifpdf
    \graphicspath{{Figures/PDF/}}
\else
    \graphicspath{{Figures/EPS/}}
\fi

\usepackage{cleveref}
\usepackage{color}

\begin{document}
\begin{frontmatter}

\title{Intra- and Inter-Frequency Brain Network Structure in Health and Schizophrenia}


\author[label1]{Felix Siebenh\"{u}hner}
\author[label2]{Shennan A. Weiss}
\author[label3]{Richard Coppola}
\author[label4,label5]{Daniel R. Weinberger}
\author[label1,label6]{Danielle S. Bassett\corref{cor1}}
\cortext[cor1]{Corresponding author email dbassett@physics.ucsb.edu}

\address[label1]{Department of Physics, University of California, Santa Barbara, CA 93106 USA}
\address[label2]{Department of Neurology, Columbia University New York, NY, 10027 USA}
\address[label3]{MEG Core Facility, National Institute of Mental Health, Bethesda, MD 20892 USA}
\address[label4]{Genes, Cognition and Psychosis Program, Clinical Brain Disorders Branch, National Institute of Mental Health, Bethesda, MD 20892 USA}
\address[label5]{Lieber Institute for Brain Development, Johns Hopkins Medical Campus, Baltimore, MD 21230 USA}
\address[label6]{Sage Center for the Study of the Mind, University of California, Santa Barbara, CA 93106 USA}

\normalsize

\newpage
\begin{abstract}
Empirical studies over the past two decades have provided support for the hypothesis that schizophrenia is characterized by altered connectivity patterns in functional brain networks. These alterations have been proposed as genetically-mediated diagnostic biomarkers and are thought to underlie altered cognitive functions such as working memory. However, the nature of this dyconnectivity remains far from understood. In this study, we perform an extensive analysis of functional connectivity patterns extracted from MEG data in 14 subjects with schizophrenia and 14 healthy controls during a 2-back working memory task. We investigate uni-, bi- and multivariate properties of sensor time series by computing wavelet entropy of and correlation between time series, and by constructing binary networks of functional connectivity both within and between classical frequency bands ($\gamma$, $\beta$, $\alpha$, and $\theta$). Networks are based on the mutual information between wavelet time series, and estimated for each trial window separately, enabling us to consider both network topology and network dynamics. We observed significant decreases in entropy in prefrontal and lateral sensor time series and significant increases in connectivity strength in the schizophrenia group in comparison to the healthy controls. We identified an inverse relationship between entropy and strength across both subjects and sensors that varied over frequency bands and was more pronounced in controls than in patients. The topological organization of connectivity was altered in schizophrenia specifically in high frequency $\gamma$ and $\beta$ band networks as well as in the $\gamma$-$\beta$ cross-frequency networks. Network topology varied over trials to a greater extent in patients than in controls, suggesting disease-associated alterations in dynamic network properties of brain function. Our results identify signatures of aberrant neurophysiological behavior in schizophrenia across uni-, bi- and multivariate scales and suggest novel candidate intermediate phenotypes in cross-frequency network architecture and in network dynamics.
\end{abstract}


\end{frontmatter}

\newpage
\section*{Introduction}
\addcontentsline{toc}{section}{Introduction}

Synchronous oscillatory brain activity is thought to form a neurophysiological network over which the brain processes information \citep{Fries2005} during cognitive processes as varied as learning and memory. Coherent neuronal communication is constrained by the underlying neurochemistry of the dopamine, GABA, glutamate and acetylcholine systems \citep{Haenschel2011,WangX2010}. Diseases like schizophrenia thought to involve such neurotransmitter systems \citep{Haenschel2011,Stephan2009,WangX2010} are therefore also often accompanied by altered oscillatory network structure.

Indeed, a hallmark of schizophrenia is the complex pattern of abnormal increases and decreases in resting state and task-based connectivity evident between large-scale brain regions \citep{Volkow1988,Weinberger1992,Friston1998}. To quantify this dysconnectivity \cite{Stephan2009} at the level of the whole brain, network theoretical tools originally developed in the social sciences have been applied to neuroimaging data. Results demonstrate complex dysconnectivity profiles of the brain's anatomy \citep{vandenHeuvel2010,Rubinov2011,Zalesky2011b} and function \citep{Bassett2008,Bassett2009a,Lynall2010,Weiss2011,Rubinov2010,Liu2008,WangX2010,Zalesky2011,Bassett2012a} in people with schizophrenia \citep{Fornito2012}.

Altered brain function in schizophrenia is unsurprisingly coupled with altered cognitive behavior. Due to the heterogeneity of the disease, some symptoms including both positive (hallucinations, delusions and thought disorder) and negative (e.g. impaired emotional response and social interaction, avolition, anhedonia) symptoms, vary strongly among patients. However, cognitive impairments in executive functions (such as the manipulation of transiently stored information  \citep{Lewis2006}) and working memory (e.g., the ability to transiently maintain and manipulate a limited amount of information in order to guide thought or behavior) are typically altered to varying degrees in virtually all patients. These cognitive impairments appear early and change little over time, and are therefore thought to lie at the core of the disease, contributing to the development of clinical symptoms \citep{Elvevag2000,Gold2004}. Both altered memory and executive planning have been linked to the neurochemistry of dopamine, GABA, glutamate and acetylcholine systems and subsequently to aberrant (NMDAR)--mediated synaptic plasticity \citep{Haenschel2011,Stephan2009,WangX2010}.

Correlations between brain network organization and cognitive or behavioral variables in schizophrenia are expected \citep{Cole2011} but quantitative evidence for such a relationship is still rare. Two recent studies have demonstrated that the organization of task-based functional brain networks in this population can be linked both to the efficacy of auditory rehabilitation efforts \citep{Weiss2011} and to working memory performance \citep{Bassett2009a,He2012,Repovs2012,WangX2010}. These findings suggest that the characterization of the dynamic network of synchronous brain activity might be a powerful tool to study cognitive impairments in schizophrenia and might provide insight into the neurobiological mechanisms of the disease. However, many of these studies have several potential limitations, such as using only a small set of network diagnostics and constructing a single functional network from a large array of data.

In this study, we provide an expanded examination of the functional brain network architecture in people with schizophrenia estimated from magnetoencephalography (MEG) data acquired during the performance of a working memory N-back task. We first examine the link between single time series variability and pairwise time series similarity, as a means of characterizing the relationship between functional activity and connectivity. To quantitatively characterize functional brain network structure, we examine 12 binary network diagnostics across a wide range of densities and employ a recently developed statistical technique (functional data analysis) to identify significant group differences. To detect time-varying network structure and to increase statistical robustness, we construct an ensemble of networks for each individual using data extracted from 66 separate trials.

Finally, in a novel extension of the analysis of functional brain network structure in the classical frequency bands of brain activity, we examine functional networks constructed from the interactions between frequency bands. Inter-frequency networks are particularly interesting in the context of our examination of the N-back task because cross-frequency interactions are thought to facilitate memory function \citep{Sauseng2008} by enabling the cross-function integration of information over spectrally distinct processing streams \citep{Palva2005}. Furthermore, evidence suggests that cross-frequency interactions are altered in schizophrenia, and the strength of these alterations has been linked to genetic risk factors for the disease \citep{Allen2011}.

Using these tools, we uncover an extensive pattern of altered network structure and network dynamics in people with schizophrenia that could potentially be used to identify intermediate phenotypes and develop diagnostic biomarkers for the disease.

\section*{Materials and Methods}											
\addcontentsline{toc}{section}{Materials and Methods}

\subsection*{Data Acquisition}
\addcontentsline{toc}{subsection}{Data Acquisition}

14 healthy volunteers and 14 people diagnosed with schizophrenia spectrum disorders (according to the Diagnostic and Statistical Manual of Mental Disorders IV criteria) took part in the Clinical Brain Disorders Branch/National Institute of Mental Health Genetic Study of Schizophrenia (National Institutes of Health Study Grant NCT 00001486, Daniel R. Weinberger, principal investigator). None of the healthy volunteers had structural magnetic resonance imaging abnormalities or history of psychiatric illness, depression, or loss of consciousness. All patients were receiving antipsychotic medication at the time of the study; none of the healthy volunteers were taking psychoactive medication.
Subjects were matched for sex (9 males, 5 females in each group) and did not differ significantly in age. The age of the healthy control group was 30 years $\pm 7.3$ (SD), and that of the patient group was 33 years $\pm 8.6$ (SD). The protocol was approved by the National Institute of Mental Health Institutional Review Board and informed consent was given in writing by all participants.

MEG data were acquired at the National Institute of Mental Health in Bethesda, MD using a 275-channel CTF system (VSM MedTech) with a sampling rate of 600 Hz. The experimental paradigm was a visual 2-back working-memory task using numerical stimuli ranging from 1 to 4 as previously described \citep{Callicott2003}. Earlier studies using the same task demonstrated abnormal behavioral and neuroimaging associations with schizophrenia \citep{Axmacher2010,Calicott2003,Pesonen2007}. Each subject performed six blocks of 11 trials (66 trials in total per subject). In each trial, a number was presented visually for 500 msec and, beginning with the third trial, the subject was asked to respond within 1,300 ms by pressing 1 of 4 buttons to indicate the identity of the number seen 2 trials previously.

\subsection*{Data Preprocessing}
\addcontentsline{toc}{subsection}{Data Preprocessing}

Using MATLAB \citep{Matlab} and FieldTrip software \citep{Fieldtrip11}, raw data were mean corrected and filtered to attenuate background low-frequency noise and line noise at 60 Hz by using a 0.3-Hz-width filter. The axial gradiometers of the CTF machine have source profiles that include information from a wide spatial range, strictly limiting the interpretation of the anatomical location of results. To retain the inherent correlation structure of a network of interacting brain regions while gaining localization specificity, we transformed the data into planar space. Using Fieldtrip, the planar transform was applied to the data by using the function \textit{ft\_megplanar}. For trial-by-trial analysis, the time series were cut into epochs of ~1.8s, beginning with the presentation of one stimulus and ending with the beginning of the next stimulus (or with the end of the block of trials in the case of the final trial). We therefore investigated a total of 66 time windows per subject.

\subsection*{Wavelet Decomposition}
\addcontentsline{toc}{subsection}{Wavelet Decomposition}
We focused our investigation on frequency-band-specific oscillations in the MEG signal by employing a wavelet analysis. Time series were resampled to 120 Hz to constrain the frequency bands of the wavelet transform to roughly conform to the classical frequency bands \cite{Bassett2009a}. Each frequency band of interest was isolated by applying the maximal overlap discrete wavelet transform to each time series \citep{Percival2000}. In line with previous work, we used the Daubechies 4 wavelet \citep{Bassett2009a}. Wavelet scale 1 (30-60 Hz) roughly corresponds to the $\gamma$ band, scale 2 (15-30 Hz) to $\beta$, scale 3 (8-15 Hz) to $\alpha$, and scale 4 (4-8 Hz) to $\theta$. Because of the relatively short trial length (1.8 s), we did not examine frequencies below 4 Hz.

\subsection*{Functional Connectivity and Complexity Estimation}
\addcontentsline{toc}{subsection}{Functional Connectivity Estimation}

Before constructing functional brain networks, we characterized each sensor's wavelet time series and the pairwise relationship between sensor time series. To quantify the complexity of the time series, we computed the Shannon wavelet entropy \citep{Shannon1948,Rosso2001,Bassett2012a}, which we describe in detail in the supplementary online materials (SoM). To investigate functional connectivity between sensors, we calculated the mutual information (MI), which is particularly appropriate for estimating interactions that encompass a narrow frequency band (e.g., a wavelet scale) \citep{David2004}.

We represent pairwise sensor MI both within and between frequency bands as functional connectivity matrices. To construct intra-frequency functional connectivity matrices, we calculated the MI between the time series in frequency band $a$ of all possible pairs of sensors to create the weight matrix $W^a$. To construct inter-frequency functional connectivity matrices, the process was similar: for each pair of frequency bands $a$ and $b$, we calculated the MI between the time series of sensor $i$ in frequency $a$ and sensor $j$ in frequency $b$ for all possible pairs of sensors $i$ and $j$. This resulted in the weight matrix $W^{ab}$. The pairwise mutual information values in both the intra- and inter-frequency matrices were normalized according to Strehl and Ghosh \citep{Strehl2002}, ensuring that the values of the elements of $W^a$ and $W^{ab}$ were in the range $[0,1]$. Through this process, we created 4 intra-frequency functional connectivity matrices ($\gamma$, $\beta$, $\alpha$, and $\theta$) and 6 inter-frequency functional connectivity matrices ($\gamma-\beta$, $\gamma-\alpha$, $\gamma-\theta$, $\beta-\alpha$, $\beta-\theta$, $\alpha-\theta$) per subject for each of the 66 trials (660 total networks per subject).

The strength of sensor $i$ was defined as the average MI between that sensor and all other sensors: $S_{i} = \frac{1}{N} \sum_{j} W_{i,j}$. The strength of a participant was defined as the average strength of all sensors: $S = \frac{1}{N} \sum_{i} S_{i}$.

\subsection*{Network Properties}

To quantitatively characterize the topology of the intra- and inter-frequency functional connectivity matrices, we construct sets of binary networks summarizing the interactions (edges) between sensors (nodes). A \emph{binary network} can be represented mathematically by an adjacency matrix $A$, whose entries $A_{ij}$ are either 0 or 1, indicating the absence or presence of an edge respectively. Binary networks can be obtained by thresholding a given functional connectivity matrix in several ways \citep{Schwarz2011,Bassett2012a}. Here we employ the technique of cumulative thresholding, a procedure in which a threshold is applied to the weighted functional connectivity matrix $W$ to retain a given percent of strongest connections. Weights (e.g., MI values) that pass this threshold were set to a value of 1 in the adjacency matrix $A$, while those that did not pass this threshold were set to a value of 0 in the adjacency matrix $A$.

The percent of nonzero elements of $A$ is also known as the network density, or network cost $\mathcal{K}$. We can employ a range of thresholds to probe the topological organization of a network over a range of network densities. Using a high threshold, we obtain a sparse matrix representing the topology of the strongest functional associations between sensors, while using a lower threshold, we obtain a denser matrix representing the topology of functional associations over a wider range of strengths. In this study, we employed a set of thresholds enabling us to examine corresponding sets of networks with densities ranging from 0 (no connections present) to 0.5 (half of the possible number of connections present) in steps of 0.01. We refer to this network density range $ 0<\mathcal{K} < 0.5$ as the cost regime of interest. Our choice to examine network organization over a large range of network density values is supported by previous studies \citep{Bullmore2009,HORSTMANN2010} that have demonstrated that the choice of threshold can have a large influence on the topological properties of the binary graph.

We quantified the organization of the binary networks using the path-length and clustering coefficient \citep{Watts1998}, global and local efficiency \citep{Latora2001,Achard2007,Bassett2009a}, betweenness centrality \citep{Freeman1977}, modularity \citep{Leicht2008,Meunier2008}, hierarchy \citep{Ravasz2003,Bassett2008}, synchronizability \citep{Barahona2002,Bassett2006a}, assortativity \citep{Newman2006,Bassett2008}, and robustness \citep{Achard2006,Lynall2010}. In addition to topological network properties, we also studied the physical measure of mean connection distance \citep{Bassett2008}, as well as a combined topophysical property Rent's exponent \citep{Bassett2010} which estimates the efficiency of the topological embedding of the network into physical space. Mathematical descriptions of these diagnostics are given in the SoM.

\subsection*{Statistical Analyses}

We examined the reliability of network properties over subjects and trials using the coefficient of variation (CV), which is defined as the standard deviation $\sigma$ of a given sample normalized by its mean value $\mu$:
\begin{equation}
 CV = \frac{\sigma}{\mu}.
\end{equation}
To test for inter-subject temporal variability in functional brain network organization, we calculated the CV for each network diagnostic, for each subject in both groups, and for all 10 intra- and inter-frequency networks. Note that we excluded the hierarchy and assortativity from this analysis because their values are close to zero, making the CV less meaningful and more prone to estimation errors from the division of numbers $\ll 1$. To test for group differences in temporal variability of network structure, we used a repeated measures ANOVA with CV values averaged over costs, group as a categorical factor, and frequency band and graph diagnostic as repeated measures.

To identify  group differences in network diagnostics, we used Functional Data Analysis (FDA). FDA enables statistical inference from sets of functions \citep{Ramsay2009} by extending the principles of statistical inference from data points to data curves. Here, in keeping with \citep{Bassett2012a}, the values of the graph properties were treated as a function of network density, and the two groups, (controls and people with schizophrenia) were compared with a non-parametric permutation test using twenty thousand permutations of group labels.

In addition to testing for group differences in network structure, it is often of interest to determine whether the structure of an empirical network is different from what one would expect in a given null model. While the development of potentially useful network null models is ongoing \cite{Zalesky2012,Bassett2012b}, here we employ benchmark Erd\"{o}s-Renyi (ER) random graphs to test whether the network topology identified in intra- and inter-frequency functional brain networks was non-random. We created an ensemble of 66 ER graphs for each network density and each of the 10 frequency bands. Note that the number of networks in the ensemble was set to be identical to the number of networks in a single subject.

All computational and statistical operations were implemented in MATLAB$^{\textregistered}$ (2007a, The MathWorks Inc., Natick, MA). Network diagnostics were estimated using a combination of in-house software, the Brain Connectivity Toolbox \citep{Rubinov2010}, and the MATLAB Boost Graph Library \citep{Matlab}. The repeated measures ANOVA was performed using freely available code \citep{AnovaRM}.

\section*{Results}												
\addcontentsline{toc}{section}{Results}
\subsection*{Working Memory Performance}

We examined the accuracy of performance in the 2-back working memory task for both the schizophrenia and control groups. People with schizophrenia had a significantly lower accuracy ($44.4 \pm 25.5$\% (STD), the median was 56.1\%) than controls ($84.2 \pm 16.7$\%): two sample t-test $t=5.15$, $p=2.23\times 10^{-5}$. These results confirm that working memory function is impaired in our cohort of people with schizophrenia, supporting an additional investigation into the patterns of brain function during task performance.

\subsection*{Time Series Variability and Co-Variability}

\begin{figure*}
\begin{center}
\includegraphics[width=0.4\textwidth]{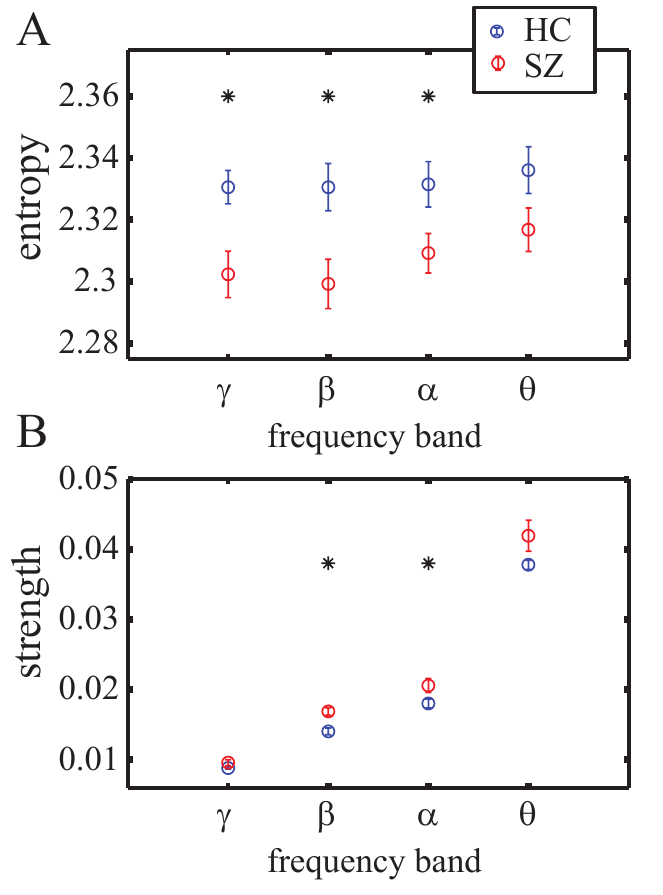}
\caption{\textbf{Group Differences in Activity and Connectivity}. Wavelet entropy (\emph{A}) and intra-frequency strength (\emph{B}), averaged over sensors, in the $\gamma$-, $\beta$-, $\alpha$- and $\theta$-bands for healthy controls and people with schizophrenia. Asterisks indicate significant group differences as measured by two-sample t-tests ($p<0.05$).
\label{fig:entropy_strength_error}}
\end{center}
\end{figure*}

\begin{figure*}
\begin{center}
\includegraphics[width=0.9\textwidth]{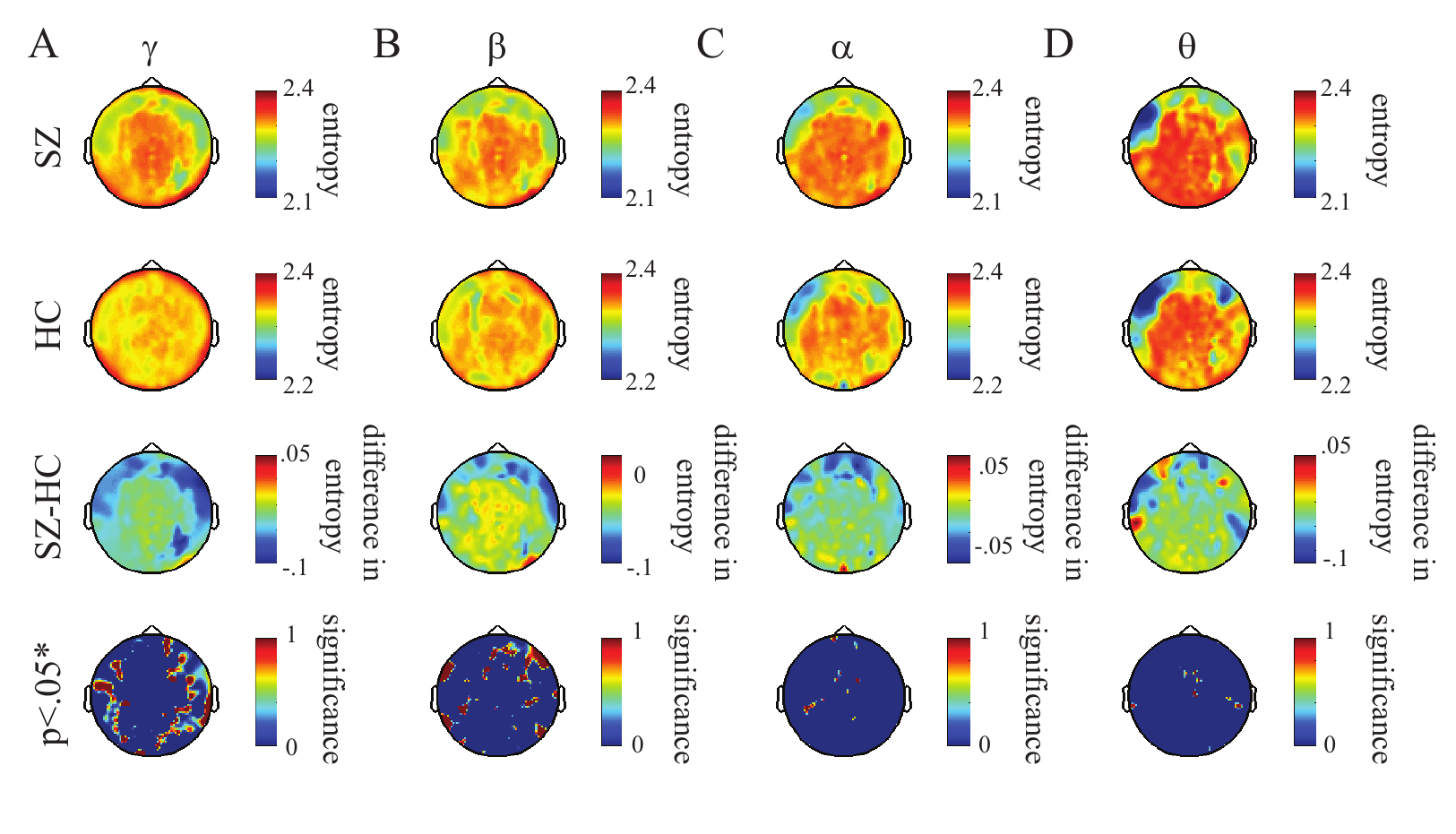}
\caption{\textbf{Anatomical Distribution of Entropy}. Topography of wavelet entropy estimated for individual sensors in the $\gamma$-, $\beta$-, $\alpha$- and $\theta$-bands (\emph{A-D} respectively), averaged over the people with schizophrenia (\emph{Top row}) and healthy controls (\emph{Second row}). Difference between groups (probands minus controls) (\emph{Third row}) were tested for significance ($p<0.05$) by permutation testing of group labels (\emph{Bottom row}).
\label{fig:entropy}}
\end{center}
\end{figure*}

\paragraph{Time Series Variability} We examined the complexity of the MEG signals in the two groups by measuring the Shannon wavelet entropy of the sensor time series in four classical frequency bands($\gamma$-, $\beta$-, $\alpha$- and $\theta$; see Methods). On average, the entropy was lower in people with schizophrenia in 3 of the 4 bands ($\gamma$, $\beta$, and $\alpha$); see Figure \ref{fig:entropy_strength_error}A. Furthermore, the distribution of entropy across sensors varied over frequency bands and between groups; see Figure \ref{fig:entropy}. The most significant group differences were identified in lateral and prefrontal areas, where entropy is lower in people with schizophrenia than in healthy controls in the high frequency $\gamma$- and $\beta$- bands.

\begin{figure*}
\begin{center}
\includegraphics[width=0.9\textwidth]{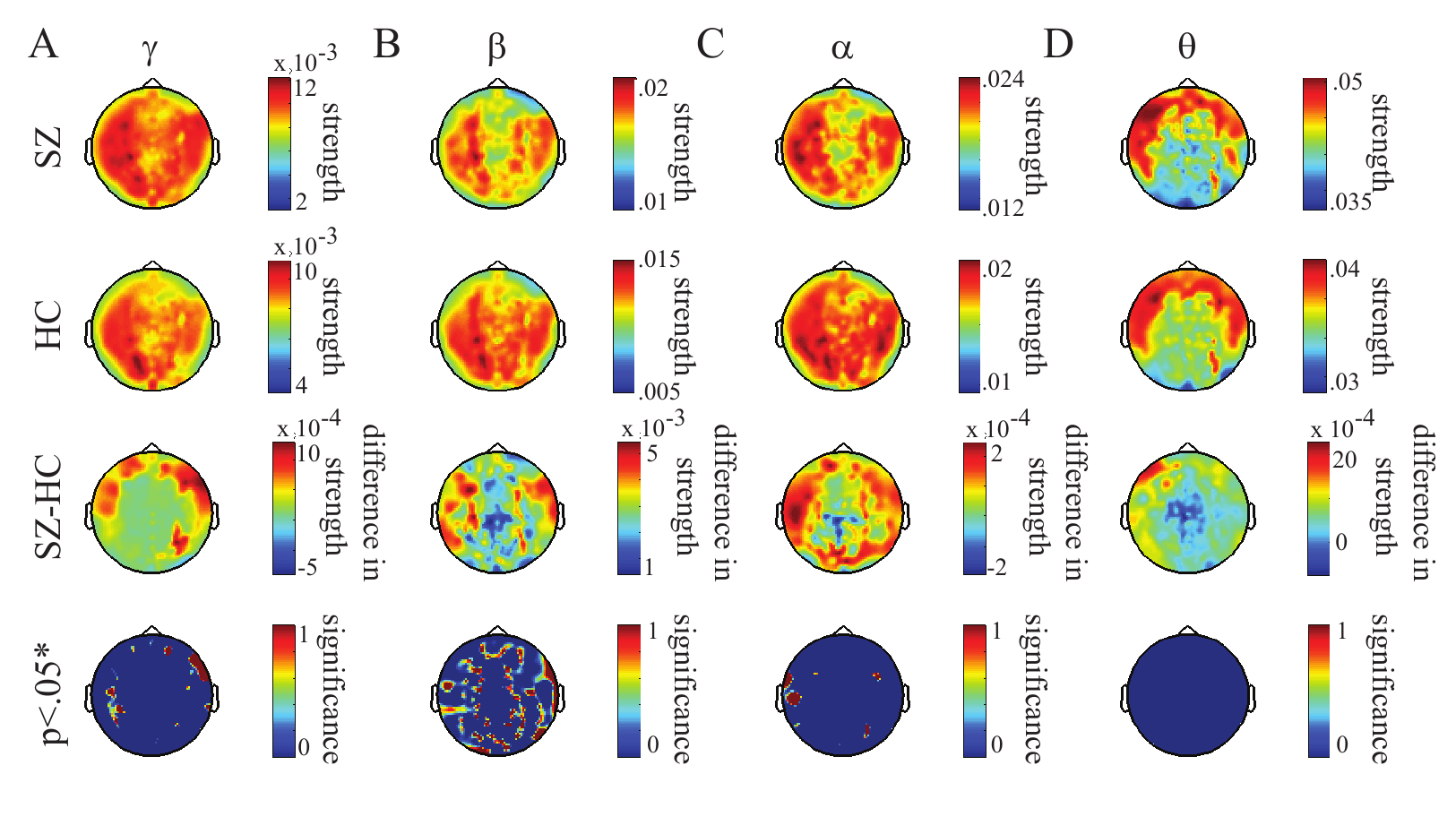}
\caption{\textbf{Anatomical Distribution of Strength}. Topography of average mutual information between individual sensors and all other sensors in the $\gamma$-, $\beta$-, $\alpha$- and $\theta$-bands (\emph{A-D} respectively), averaged over the people with schizophrenia (\emph{Top row}) and healthy controls (\emph{Second row}). Difference between groups (probands minus controls) (\emph{Third row}) were tested for significance ($p<0.05$) by permutation testing (\emph{Bottom row}).
\label{fig:strength}}
\end{center}
\end{figure*}

\paragraph{Time Series Co-Variability} The co-variability of sensor time series is thought to be a measurement of synchronous oscillatory neuronal activity and therefore potentially a proxy for large-scale communication between brain regions. We estimated the mutual information between sensor time series in the same four frequency bands ($\gamma$-, $\beta$-, $\alpha$- and $\theta$; see Methods). The strength of a sensor was defined as the average mutual information between that sensor's time series and the time series of all other sensors. Strength displayed frequency-dependent variation, being smallest in the highest frequency band ($\gamma$) and largest in the lowest frequency band ($\theta$) for both groups; see Figure \ref{fig:entropy_strength_error}. On average, the strength was higher in people with schizophrenia in 3 of the 4 bands ($\gamma$, $\alpha$, and $\theta$). Furthermore, the distribution of strength across sensors varied over frequency bands and between groups; see Figure \ref{fig:strength}. We found that strength was larger in people with schizophrenia, particularly in prefrontal sensors in the $\gamma$-band and in prefrontal and other lateral sensors in the $\beta$- band.

\paragraph{Relationship Between Entropy and Strength} Recent work has suggested a potential and intuitive link between time series variability (e.g., entropy) and time series co-variability (e.g., correlation or mutual information) in fMRI data over subjects \citep{Zalesky2011} and over brain regions \citep{Bassett2012a}. Here we examine this potential relationship in a different imaging modality (MEG) both across frequency bands and between groups. In Figure \ref{fig:str_ent}, we show the strength and entropy of each individual, averaged over sensors and trials. Importantly, we find that correlations between the two measurements at the inter-subject level are frequency-dependent. Both groups display a significant inverse relationship between entropy and strength in the $\beta$-band, suggesting that individuals with high temporal variability in brain function have lower temporal co-variability. This relationship also appears to hold for the controls in the other frequency bands. In Figure \ref{fig:str_ent_ch}, we show the strength and entropy for each sensor (rather than each individual), averaged over trials and over individuals within a group. Again, we identify a strong inverse relationship between entropy and strength, which in this case is strongest in the $\theta$-band. Of note, the inhomogeneous distribution of values characterized by a sparse low entropy tail is consistent with data reported in \citep{Bassett2012a}, and potentially indicates variation in the roles of brain areas in information processing.

\begin{figure*}
\begin{center}
\includegraphics[width=0.9\textwidth]{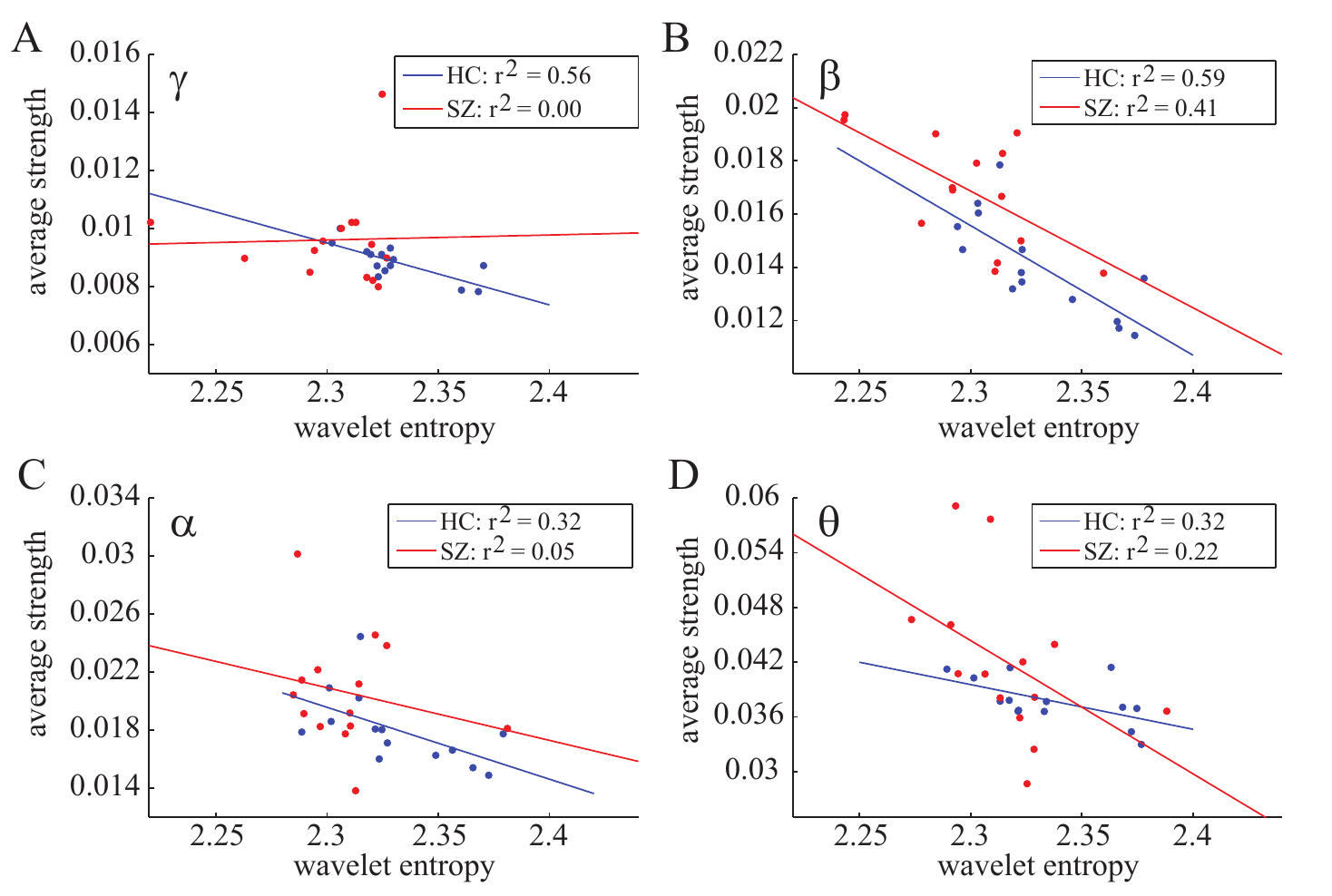}
\caption{\textbf{Entropy and Strength: Inter-Subject Level.}
Scatterplots of the strength of connectivity and the complexity of the time series, as measured by wavelet entropy, for $\gamma$-, $\beta$-, $\alpha$- and $\theta$-bands. Single data points represent values for each individual averaged over trials and sensors. Red markers denote subjects with schizophrenia spectrum diagnosis; blue markers denote healthy subjects. Lines indicate best linear fits for the two groups separately (red, SZ; and blue, HC) and we provide $r^2$ values as indicators of goodness of fit. Similar scatterplots that code each experimental block separately are provided in Figure S1 in the SoM.
\label{fig:str_ent}}
\end{center}
\end{figure*}

\begin{figure*}
\begin{center}
\includegraphics[width=0.9\textwidth]{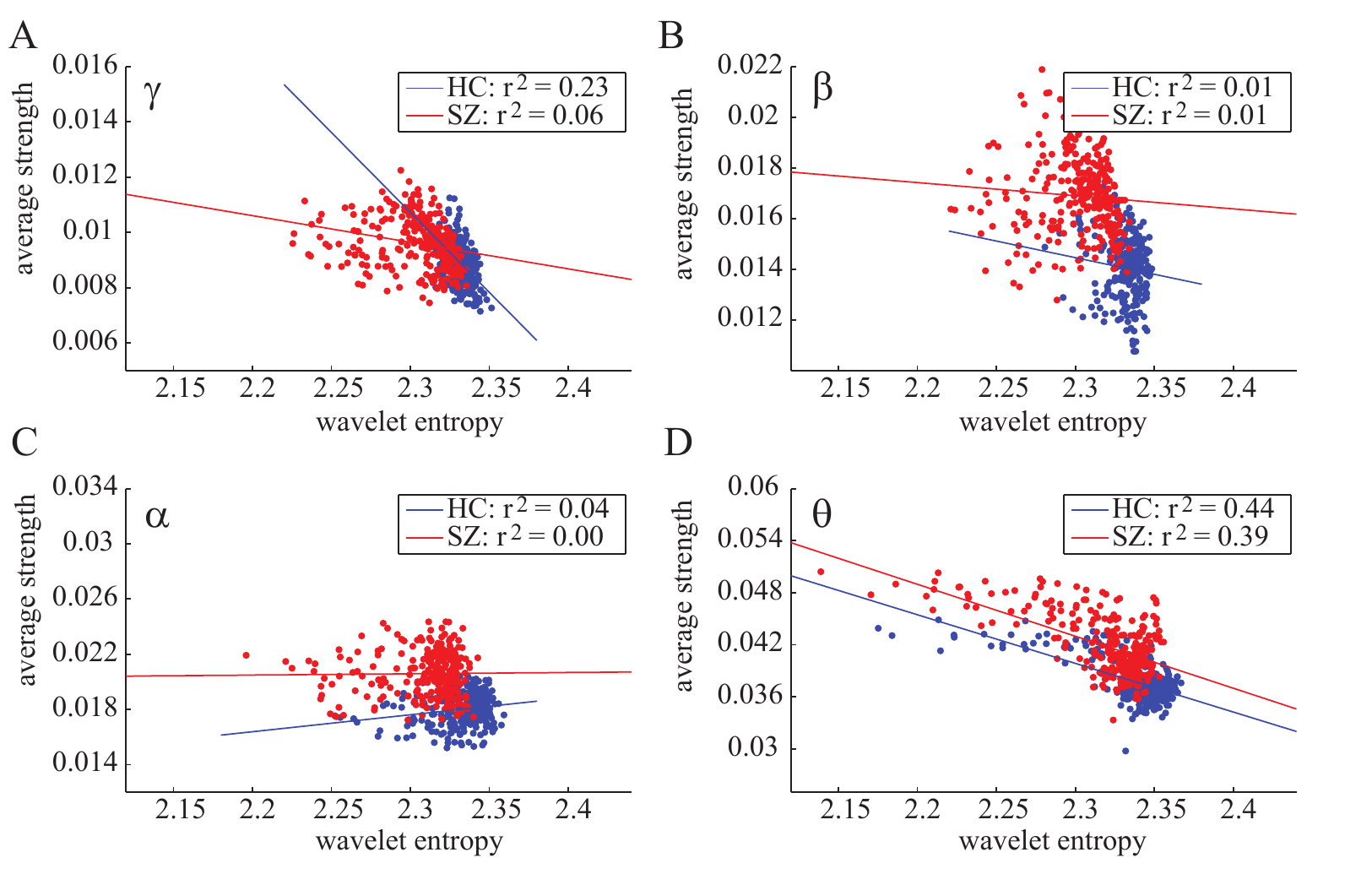}
\caption{\textbf{Entropy and Strength: Sensor Level.}
Scatterplots of the strength and wavelet entropy in $\gamma$-, $\beta$-, $\alpha$- and $\theta$-bands. Single data points represent values for each sensor averaged over trials and individuals. Red markers denote data from the SZ group; blue markers denote data from the HC group. Lines indicate linear fits for the two groups separately (red, SZ; and blue, HC) and we provide $r^2$ values as indicators of goodness of fit.
\label{fig:str_ent_ch}}
\end{center}
\end{figure*}

\subsection*{Network Structure}

Given the group differences in connectivity strength identified in the previous section, we next asked whether the patterns of connectivity between sensors are altered in schizophrenia. We characterize these patterns using binary network diagnostics. See Methods for details on binary network construction and the SoM for mathematical definitions of network diagnostics.

\begin{figure*}
\begin{center}
\includegraphics[width=0.4\textwidth]{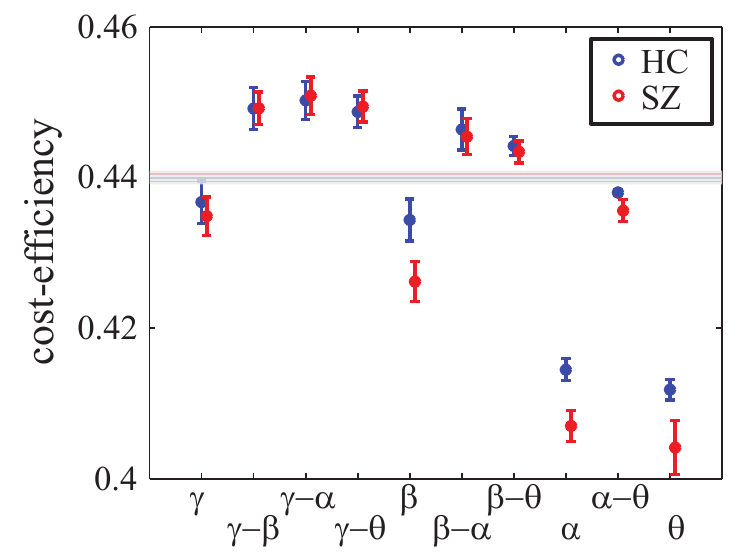}
\caption{\textbf{Cost-Efficiency of Functional Networks in Health and Disease.}
\emph{(A)} Cost-efficiency in the $\gamma$-, $\beta$-, $\alpha$-, $\theta$-, and cross-frequency band networks for healthy controls (blue) and people with schizophrenia (red). Error bars indicate the standard error over subjects. The gray shaded line indicates the expected values of cost-efficiency for an ensemble of random (ER) graphs (see Methods).
\label{fig:CE}}
\end{center}
\end{figure*}

\paragraph{Cost-Efficiency}
We constructed binary graph diagnostics as a function of network density using a cumulative thresholding technique (see Methods). A simple network diagnostic that collapses such a curve into a single value is the cost-efficiency, which is defined as the maximum of the efficiency-minus-cost curve \citep{Achard2007,Bassett2009a,Fornito2011} (see SoM). In Figure \ref{fig:CE}A, we show the cost-efficiency of both intra- and inter-frequency networks for both groups. As expected from previous work \citep{Bassett2009a}, we found that cost-efficiency values decreased with increasing frequency. Importantly, intra-frequency networks demonstrated consistently lower cost-efficiency than inter-frequency networks. Group differences were only evident in intra-frequency networks (e.g., the $\beta$, $\alpha$ and $\theta$ networks), with cost-efficiency being higher in the control group than in the patient group. Inter-frequency networks, while not demonstrating a group difference, did show a significantly higher cost-efficiency than expected in an ensemble of Erd\"{o}s-Reny\`{\i} random graphs (see Methods), suggesting the presence of non-random structure in cross-frequency interactions.

\paragraph{Binary Network Organization}

\begin{figure*}
\begin{center}
\includegraphics[width=0.9\textwidth]{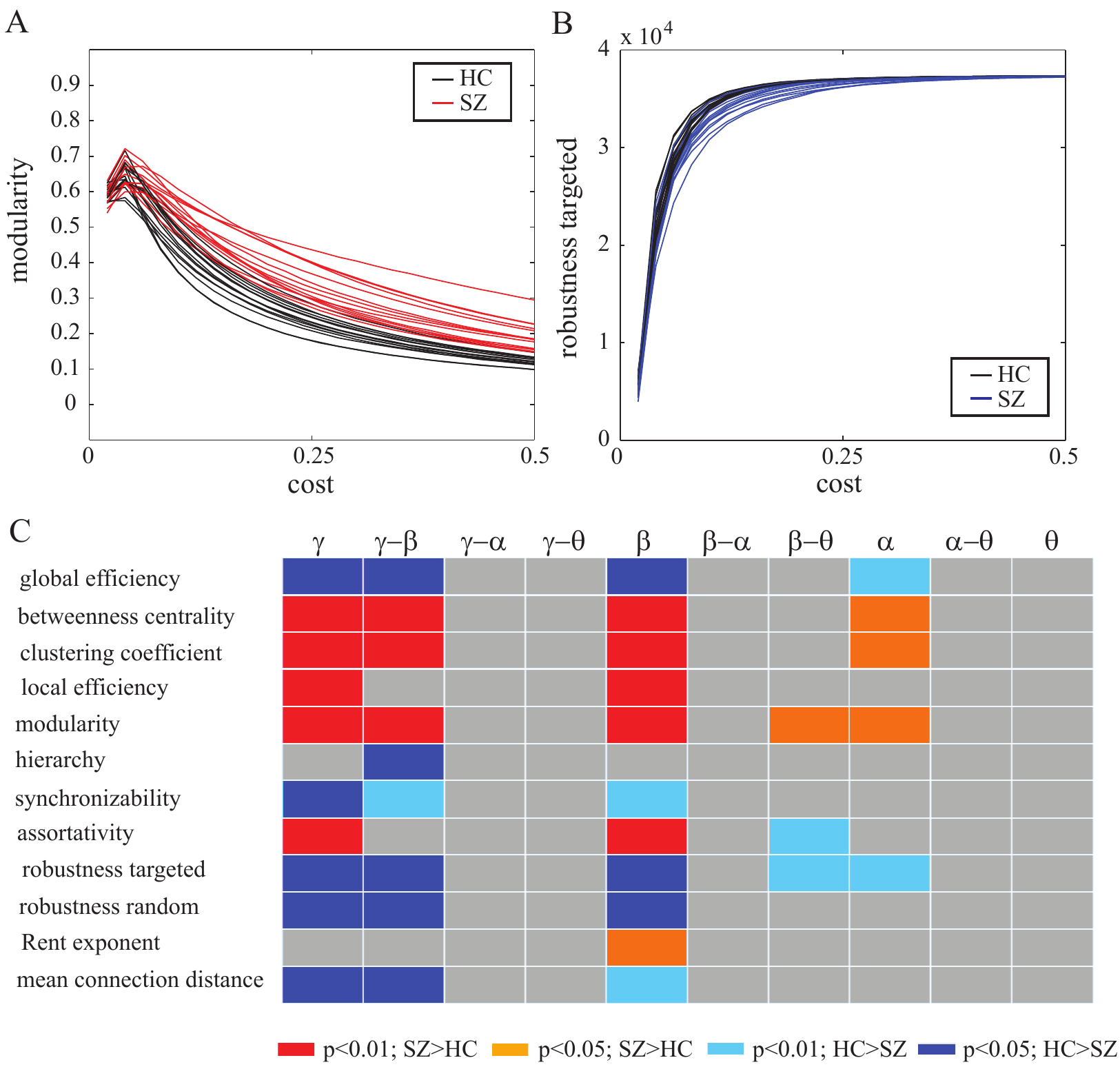}
\caption{\textbf{Binary Network Organization in Health and Disease}
\emph{(A-B)} Example binary network diagnostic curves as a function of threshold: the modularity for controls (black) and people with schizophrenia (red) \emph{(A)}, and the robustness to targeted attack for controls (black) and people with schizophrenia (blue) for $\gamma$-band networks. Individual curves indicate average values for each individual over the 66 trial-specific networks. \emph{(C)} Significant group differences in graph diagnostic versus cost curves for 12 graph diagnostics (y-axis) in both intra- and inter-frequency bands (x-axis). Warm colors indicate that the diagnostics values were higher in people with schizophrenia than in healthy controls (red, $p<0.01$; orange, $p<0.05$ uncorrected); cool colors indicate that the values were higher in healthy controls than in people with schizophrenia (dark blue, $p<0.01$; light blue, $p<0.05$ uncorrected). See SoM for the full graph diagnostic versus cost curves for all subjects, frequency bands, and diagnostics.
\label{fig:bin_metrics}}
\end{center}
\end{figure*}

While cost-efficiency has the advantage of collapsing a binary-diagnostic versus cost curve into a single value, it is also of interest to examine the shape of these curves for the other diagnostics. We examined 12 binary graph diagnostics as a function of cost for both groups and all 10 frequency bands \citep{Bassett2011a}; see Methods and SoM. We separated diagnostics into those that showed higher values in people with schizophrenia (e.g., see Figure \ref{fig:bin_metrics}A) and those that showed higher values in healthy controls (e.g., see Figure \ref{fig:bin_metrics}B). We found that the majority of significant group differences were located in the $\gamma$ and $\beta$ intra-frequency networks and in the $\gamma$-$\beta$ cross-frequency networks; see Figure \ref{fig:bin_metrics}C. Note that the $\alpha$-band also showed group differences for 5 out of the 12 diagnostics. People with schizophrenia displayed higher global and local efficiency, betweenness centrality, clustering coefficient, modularity, and assortativity and displayed lower hierarchy, synchronizability, mean connection distance and robustness to both targeted and random attack. The identification of group differences in both topological and physical network properties suggests that disease-associated changes in brain network organization might be linked to multiple developmental mechanisms. Indeed, recent theoretical work has suggested that altered constraints on both information efficiency (a topological property) and metabolic cost (a physical property) can lead to schizophrenia-like changes in network architectures \cite{Vertes2012}.

\paragraph{Variability of Binary Network Structure}

Thus far we have reported results for each individual derived from networks constructed from 66 trial blocks. Here we ask whether network organization is variable over trial blocks in the two groups. To quantify temporal variability in network structure, we computed the coefficient of variation (CV) for network diagnostics over trial blocks for each individual in all 10 frequency bands.

In Figure \ref{fig:CV}, we show the CV for the $\gamma$ and $\beta$ intra-frequency and the $\gamma-\beta$ inter-frequency networks, averaged over costs (for similar results in other frequency bands, see the SoM). We observe that the CV varies over binary graph diagnostics in a similar way across frequency bands. We also note that in almost all cases, the temporal variability appears to be larger for people with schizophrenia. To quantitatively test this observation, we used a Repeated Measures ANOVA (see Methods for details and Table 1 for results). The main effect of group ($F=6.24$, $p=0.02$) confirmed that the networks derived from people with schizophrenia varied more over time than did those derived from healthy controls, suggesting a fundamental alteration in the dynamics of functional brain networks in schizophrenia.

\begin{figure*}
\begin{center}
\includegraphics[width=0.5\textwidth]{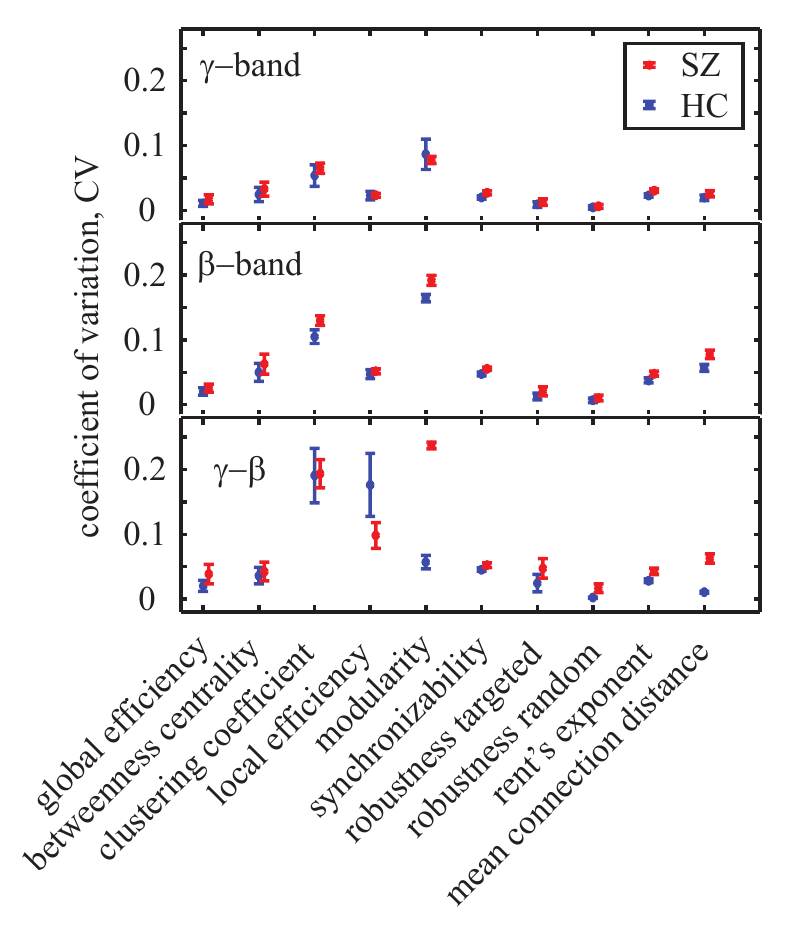}
\caption{\textbf{Temporal Variability of Network Diagnostics} for binary network diagnostics in the $\gamma$ and $\beta$ intra-frequency networks and in the $\gamma-\beta$ inter-frequency networks. CV values indicate temporal variability over trials for healthy (blue) and schizophrenic subjects (red), averaged over the entire range of cost values. Error bars indicate the mean squared error over subjects and costs. Values and error bars for synchronizability are scaled down by a factor of 10 for visualization purposes. Results for other frequency bands can be found in the SoM.
\label{fig:CV}}
\end{center}
\end{figure*}

\begin{table}[htbp]
\caption{\textbf{Results for a Repeated Measures ANOVA of Temporal Variability in Network Topology}, measured using the CV.}
\begin{center}
\begin{tabular}{|l|r|r|r|r|r|}
\hline
 & \multicolumn{1}{l|}{SSE} & \multicolumn{1}{l|}{DF} & \multicolumn{1}{l|}{MSE} & \multicolumn{1}{l|}{F} & \multicolumn{1}{l|}{p} \\ \hline
Group & 0.04 & 1 & 0.04 & 6.24 & 0.02 \\ \hline
Frequency Band & 0.41 & 9 & 0.05 & 52.6 & 0 \\ \hline
Diagnostic & 6.24 & 9 & 0.69 & 309.74 & 0 \\ \hline
Frequency Band*Diagnostic & 1.63 & 81 & 0.02 & 41.83 & 0 \\ \hline
Frequency Band*Group & 0.04 & 9 & 0 & 4.52 & $1.68e^{-5}$ \\ \hline
Diagnostic*Group & 0.11 & 9 & 0.01 & 5.27 & $1.52e^{-6}$ \\ \hline
Frequency Band*Diagnostic*Group & 0.11 & 81 & 0 & 2.74 & $4.64e^{-14}$ \\ \hline
\end{tabular}
\end{center}
\label{anova}
\end{table}

\section*{Discussion}
\addcontentsline{toc}{section}{Discussion}

We have examined temporal characteristics of MEG data acquired from people with schizophrenia and controls during a working memory task. Our approach spanned several distinct levels, including that of the individual sensor time series (univariate), the co-variability between time series (bivariate), and the patterns of co-variability across sensors using binary network diagnostics (multivariate). We identified disease-associated changes in brain function at each level. People with schizophrenia displayed lower time series entropy, higher strength of co-variability between time series, and an extensive pattern of altered topological organization in binary sensor networks. Importantly, network properties of cross-frequency associations between time series in the $\gamma$- and $\beta$-bands differed between groups, uncovering a novel feature of dysconnectivity. Finally, the temporal variability of brain network architecture in people with schizophrenia was significantly higher than that in healthy controls, a phenomenon suggestive of decreased dynamic constraints on brain function.

\subsection*{Identification of Intermediate Phenotypes}
The mechanisms of psychotic symptoms in schizophrenia, their genetic underpinnings and their brain signatures, are far from understood. The interpretations of neuroimaging phenomena and related conclusions regarding neurophysiological mechanisms of the disease, are greatly hampered by the confounding effects of medication and disease heterogeneity associated with epiphenomena related to smoking and chronicity, as well as the heterogeneity of the mental state of ill subjects during these MEG procedures. Thus, it is impossible to conclude that our findings represent primary disease phenomena rather than epiphenomena related to secondary factors that are associated with the state of illness. These difficulties can be overcome to some degree by leveraging the genetic similarity between people with schizophrenia and their unaffected siblings and testing for similar imaging phenomena in both groups.

An important goal of this study was to perform an extensive assessment of alterations in brain functional architecture in schizophrenia to identify candidate diagnostics for intermediate phenotypes of the disease \cite{Fornito2012b}. Intermediate phenotypes -- observable characteristics that show tiered values in patients, their siblings, and controls -- represent a uniquely powerful tool for the quantification of genetic risk mechanisms for psychiatric disease \citep{Lindenberg2006}. In particular, phenotypes derived from human neuroimaging have intriguing potential to provide concrete insight into the neurochemical pathologies \citep{Hirvonen2011} that might directly underlie the behavioral symptoms of psychosis \citep{Tan2011}.

Our multiscale exploration of illness-associated physiologic phenomena and their potential to represent intermediate phenotypes has been guided by evidence for genetic regulation of both brain activity variability \citep{Winterer2004} and functional connectivity architecture \citep{Fornito2011}. Temporal variability of brain activity specifically during memory tasks has previously been identified as an intermediate phenotype of schizophrenia \citep{Winterer2004}, validating the use of the $N$-back task in our study. Further examining the patterns of functional connectivity supporting memory function is critical given recent evidence suggesting that functional connectivity can be more sensitive to genetics \citep{Esslinger2009} and to schizophrenia \citep{Bassett2012a,Breakspear2003} than the simple activity of brain regions alone.

Our results suggest that network organization in the higher frequency bands ($\gamma$ and $\beta$) is potentially a better candidate for an intermediate phenotype than that in lower frequency bands during the $N$-back task. Importantly, in this study, we have also introduced two other potential candidates for intermediate neuroimaging phenotypes based on signatures of dynamic sensor-sensor communication: 1) network properties of whole-brain inter-frequency connectivity and 2) dynamic changes in network configuration during task performance. We find that the $\gamma$-$\beta$ networks display significant group differences in organization, similar to the patterns seen in the high frequency bands alone. Furthermore, the variability in this network structure over the duration of task performance is significantly higher in people with schizophrenia than in controls, suggesting that dynamic reconfiguration properties of the healthy brain might be affected in the disease. Taken together, our results lay the groundwork for a directed examination of the genetic underpinnings of synchronous oscillatory brain activity in probands and their unaffected siblings. Importantly, as deficits in cognitive function such as working memory are thought to be fundamental and possibly causal in schizophrenia \citep{Elvevag2000,Gold2004}, such studies might eventually lead to improved detection and treatment of the disease.

\subsection*{Altered Time Series Variability}
Signal variability or noise is a characteristic feature of the cortical system \citep{Faisal2008} and is thought to facilitate the exploration of functional network configurations necessary for healthy cognitive function \citep{Deco2011}. Increased variability of stimulus-induced prefrontal electromagnetic activity has been observed in both people with schizophrenia and their healthy siblings \citep{Winterer2000,Winterer2004,Winterer2006}, suggesting that cortical noise might be a genetic biomarker for the disease. Indeed, distributed patterns of both increased and decreased signal variability differentiate people with predominantly positive or negative symptom profiles \citep{Raghavendra2009,John2009}. Furthermore, signal variability is behaviorally relevant, having been linked to task accuracy in both probands and healthy controls \citep{Winterer2004,Winterer2006,Krishnan2005,Kim2008}.

Our results support the notion that signal variability is affected in schizophrenia. By employing the simple measure of time series entropy, we showed that signal variability was decreased in probands specifically in prefrontal and lateral sensors in high frequency bands ($\gamma$ and $\beta$). However, it is important to note that the Shannon entropy characterizes properties of the distribution of values in the time series rather than the temporal evolution of the signal. Alternative measures of temporal signal variability, such as the multi-scale entropy \citep{Takahashi2010}, the Renyi number \citep{Gonzalez2000}, the Lyapunov exponent \citep{Xie2008}, the fractal dimension \citep{Rubinov2009b,Raghavendra2009}, and the Hurst exponent \citep{Bullmore2001}, might provide greater or lesser sensitivity to disease state \citep{Sabeti2009} or to disease-associated alterations in lower frequency bands.

\subsection*{Activity and Connectivity}

Intuitively, one might imagine that the complexity of a region's time series would have some bearing on the strength of connectivity between that region and other regions. For example, a region with a more noisy signal might be less likely to show strong associations with other regions than a region with an ordered, strong oscillatory signal.

Two recent fMRI studies have supported the view of a strong positive correlation between time series variability and co-variability \citep{Bassett2012a,Zalesky2011}. However, our results suggest a more complex relationship in MEG data that is modulated by both frequency and disease. At the subject level, an inverse correlation between entropy and strength was evident in the $\beta$ band for both groups, and also in the $\gamma$, $\alpha$ and $\theta$ bands for the control group. Similarly at the sensor level, the entropy and strength were again inversely correlated, but in this case the effect was largest in the $\theta$ band. Importantly, the sensor-level analysis uncovered a strong heterogeneity in the distribution of entropy across sensors, complicating the statistical estimate of the entropy-strength relationship.

The consistent inverse relationship between entropy and strength identified in this study is at odds with the positive relationship identified in previous fMRI studies \citep{Bassett2012a,Zalesky2011}. In our case, high entropy sensors tend to have weak connectivity to other sensors, while in the two previous fMRI studies, regions or individuals characterized by high entropy tended to also show strong connectivity. However, the divergence between results from MEG and fMRI studies might not be that surprising, as the two imaging modalities measure inherently different properties of neurophysiological activity (blood flow and magnetic flux) at different temporal resolutions. The two imaging modalities also appear to capture signatures of brain communication differently; for example, increases in the BOLD signal have been linked to both synchronization \emph{and} desynchronization in MEG oscillations \citep{Winterer2007}. Further work in both imaging modalities will be necessary to gain a greater intuition for the role of activity-connectivity relationships in cognitive function.

\subsection*{Network Structure of Connectivity}

Both binary and weighted network analyses each have their specific advantages and disadvantages and, when applied to the same data, can yield diverging results \citep{HORSTMANN2010,Cheng2012}. Weighted networks suffer from issues of normalization, which are particularly important in the context of comparing groups with different average strength (as is the case here in our study), and further require the careful identification of appropriate null models \citep{Rubinov2010,Zalesky2012,Bassett2012b}. Binary networks on the other hand rely heavily on the choice of thresholds, and neglect information contained in the weights of connections. However, binary networks are mathematically simpler \citep{Rubinov2010} and their disadvantages can be at least partially overcome by examining diagnostic properties over a range of network densities. In this paper, we focus on binary networks and examine their properties over a wide and dense cost range. To identify group differences, we employ functional data analysis \citep{Bassett2012a} and find that people with schizophrenia display higher local efficiency, betweenness centrality, clustering coefficient, modularity, and assortativity and display lower global efficiency, hierarchy, synchronizability, mean connection distance and robustness to both targeted and random attack than healthy controls in high frequency band networks.

In comparing our findings to the literature, we note that results of connectivity analyses seem to differ between imaging modalities. Network organization in schizophrenia has been examined using a variety of structural (sMRI, DTI \citep{Bassett2008,Zalesky2011,Wang2012}) and functional (fMRI, EEG, and MEG) neuroimaging techniques \citep{Rubinov2009,Bassett2012a,Liu2008,Wang2010,Lynall2010,He2012,AlexanderBloch2010,AlexanderBloch2012}.
While schizophrenia-related alterations in network structure are reported across all imaging modalities and experimental paradigms, the patterns of these differences do not always converge. One relatively consistent finding is that global network efficiency is lower in schizophrenia subjects for networks derived from DTI \citep{Zalesky2011,Wang2012}, resting state fMRI data \citep{Liu2008,Wang2010}, and our current MEG study. However, the clustering coefficient has been reported to be lower in schizophrenia for several resting state fMRI and EEG studies \citep{Rubinov2009,Lynall2010,Liu2008} but is higher in our investigation of task-based MEG in one inter- and several intra-frequency networks. While this divergence of results could be due to the different imaging modalities, another possible interpretation might be that the direction of network organizational changes in schizophrenia could depend on brain state (e.g., rest or task). Such a view is supported by recent work demonstrating that network organization changes as a function of cognitive load \citep{He2012,Kitzbichler2011} and learning \citep{Bassett2011b}.

The complex landscape of network changes in schizophrenia demonstrates the need for multimodal imaging studies on the same group of subjects, using the same methods of network construction, and employing the same graph diagnostics. Complementary theoretical work, using for example neural mass models \citep{David2003,Moran2007,Stephan2006b}, could potentially explain mechanistically why one might expect different disease-related changes in network structure in different functional or structural imaging modalities.

\subsection*{Dynamic Network Variability}

We explored the evolution of functional connectivity during the performance of a working memory task. This exploration was guided by theoretical work proposing the critical role of functional network reconfigurations in healthy cognitive function \cite{Deco2011} and by empirical work demonstrating the importance of reconfiguration flexibility for cognitive phenomena like learning \cite{Bassett2011b}. We identified greater temporal network variability on average in people with schizophrenia than in healthy controls. This finding is consistent with theoretical formulations of dysconnectivity in schizophrenia that suggest a temporal disorganization of sequentially expressed dynamical states \citep{Breakspear2005}. Indeed, from a theoretical point of view, the inability of the dynamic brain network to sustain the organization necessary for healthy cognitive function might empirically result in the fragmentation of cortical and subcortical networks seen in schizophrenia \citep{vandenBerg2012}. Empirically, our findings are also consistent with a recent EEG study by Schoen and colleagues \citep{Schoen2011} demonstrating an increased entropy of connections in schizophrenia specifically in the high frequency $\gamma$ band. Together, these results point toward altered temporal trajectories of whole-brain cortical function potentially due to inadequate or altered constraints on brain dynamics in schizophrenia.

\subsection*{Methodological Limitations and Considerations}
\addcontentsline{toc}{subsection}{Methodological Limitations}

In addition to intra-frequency investigations, we have constructed inter-frequency networks to determine whether cross-frequency communication patterns are altered in schizophrenia. Cross-frequency interactions are thought to facilitate the cross-function integration of information over spectrally distinct processing streams \citep{Palva2005} and might be critical for memory function \citep{Sauseng2008}. Phase-amplitude coupling and similar methods \citep{Canolty2010,Axmacher2010,Allen2011} have been used to quantify these interactions, but the theoretical foundations of these methods are still not well understood \citep{Vicente2011,Muthukumaraswamy2011}. We have chosen to use a relatively simple nonlinear measure of interaction -- the mutual information between times series extracted from different wavelet bands -- to quantify statistical associations between frequencies that could be studied from a network perspective. Although outside of the scope of the present paper, it would be interesting in future to examine the effects of alternative estimates of these interactions on network structure.

The data used for this study was acquired using a CTF machine, whose axial gradiometers have source profiles that include information from a wide spatial range, limiting the potential for anatomical interpretations. Commonly used source localization techniques allow for greater confidence in anatomical localization but simultaneously change the correlation structure between time series. Instead, to retain the inherent network correlation structure and increase the localization specificity, we transformed the data into planar space \citep{Bassett2009a,Weiss2011}.

In the present study, we did not observe significant correlations between the single-valued diagnostics (entropy, strength, cost-efficiency and variability) and the accuracy of task performance, for which our small sample size might be a factor. A more extensive examination of the relationship between behavioral variables and network diagnostics is outside of the scope of this study, whose focus was primarily to identify alterations in network structure and dynamics in schizophrenia. In this study, we have reported results at the statistical threshold of $p<0.05$ uncorrected for multiple comparisons. Our choice is justified given the exploratory nature of the work, and is moreover supported by the fact that it is not yet known how to correct for comparisons across multiple network diagnostics. The interdependence of these measures is not well understood, but preliminary evidence suggests the presence of non-trivial correlations between them \citep{Lynall2010,Bassett2012a}, such that strict multiple comparisons correction would not be appropriate. Furthermore, because all of the people with schizophrenia included in this study were on medication, most were smokers, and their performance on the task was significantly worse than the performance of the normal subjects, we can not determine whether our findings are driven by the disease, by associated epiphenomena, as an effect rather than a cause of the poor task performance (e.g. altered attention, effort, distraction) or by medication or a combination of these. To examine the effects of these various epiphenomena, it would be important to perform a follow-up study in unaffected siblings, for which our study gives important guiding information.

\subsection*{Conclusion}
\addcontentsline{toc}{subsection}{Conclusion}

Recent advances in physics and mathematics have provided unique, robust quantitative network methods to examine the structure and organization of whole-brain functional connectivity. Mounting evidence from a plethora of imaging modalities, cognitive states, and diseases underscores the utility of network theory in capturing previously uninvestigated variations in large-scale brain function and its alteration in disease states \citep{Bassett2009b}. However, the interpretation of these findings is complicated by the fact that network signatures of schizophrenia are not consistent across imaging modalities and analytic methods and only provide indirect information about the actual mechanisms responsible for the results. In this study, we have introduced several new methods of studying dynamic properties of brain networks in schizophrenia, including cross-frequency networks and temporal variation in network structure, and we hope that they will contribute to such advancements.

\section*{Acknowledgments}
We thank Jean M. Carlson, Karl Doron, and Scott T. Grafton for useful discussions. This work was supported by the Errett Fisher Foundation, the Templeton Foundation, David and Lucile Packard Foundation, PHS Grant
NS44393, Sage Center for the Study of the Mind, and the Institute for Collaborative Biotechnologies through contract no. W911NF-09-D-0001 from the U.S. Army Research Office. We acknowledge support from the Center for Scientific Computing at the CNSI and MRL: an NSF MRSEC (DMR-1121053) and NSF CNS-0960316. The patients and normal controls  were studied as part of the Clinical Brain Disorders Branch Sibling study (D. R. Weinberger PI), an investigation of the biological components of genetic risk for schizophrenia, with direct funding of the Weinberger Lab and the NIMH MEG core facility (R. Coppola). The content is solely the responsibility of the authors and does not necessarily represent the official views of any of the funding agencies.

\newpage
\newpage
\newpage

\section*{Supplementary Material for Intra- and Inter-Frequency Brain Network Structure in Health and Schizophrenia}

This supplementary materials section includes the following:
\begin{itemize}
\item Mathematical definitions for network diagnostics employed in the present analysis
\item Supplementary Figure 1: Entropy and Strength
\item Supplementary Figure 2: Network Diagnostics: Part I
\item Supplementary Figure 3: Network Diagnostics: Part II
\item Supplementary Figure 4: Variability of Network Diagnostics: Part II
\end{itemize}

\newpage
\subsection*{Mathematical Definitions}											
\addcontentsline{toc}{subsection}{Mathematical Definitions}

\subsubsection*{Time Series Diagnostics}
\addcontentsline{toc}{subsubsection}{Time Series Diagnostics}

\emph{Wavelet Entropy}: Entropy, as defined by Shannon \citep{Shannon1948} is a simple, time-independent measure of the degree of order/disorder of the signal and has been applied to measure the univariate complexity of signals obtained in neuroimaging \citep{Rosso2001,Bassett2012a}. Here, we calculated wavelet entropy with the MATLAB function \textit{wentropy.m}:
\begin{equation}
E(s) = - \sum_{i} s_{i}^{2} \mathrm{log}(s_{i}^{2})
\end{equation}
where $s$ is the signal of a single region in a given individual and $s_{i}$ are the coefficients of $s$ in the orthonormal wavelet basis.

\subsubsection*{Network Diagnostics}
\addcontentsline{toc}{subsubsection}{Network Diagnostics}

A network is composed of units (\emph{nodes}) and connections between those units (\emph{edges}). The degree $k_{i}$ of node $i$ is defined as the number of edges emanating from node $i$.

\emph{Clustering coefficient}
The clustering coefficient $C$ is defined by supposing that a node $i$ has $k_{i}$ neighbors, so a maximum of $k_{i} (k_{i} - 1)/2$ edges can exist between these neighbors \citep{Watts1998}. The local clustering coefficient $C_{i}$ is the fraction of these possible edges that actually exist:
\begin{equation}
	C_i=\frac{\sum_{mj} A_{mj} A_{im} A_{ij}} {k_i(k_i-1)}\,. \label{eq:Cc}
\end{equation}
The clustering coefficient $C$ of an entire network is then defined as the mean of $C_{i}$ over all nodes $i$.

\emph{Hierarchy}
A sense of hierarchical structure of the network can be characterized by the coefficient $\beta$, which is a parameter quantifying the putative power law relationship between the clustering coefficient $C_{i}$ and the degree $k_{i}$ of all nodes in the network \citep{Ravasz2003}:
\begin{equation}\label{hierarchy}
C_{i} \sim k_{i}^{- \beta}.
\end{equation}
Pragmatically, we estimate $\beta$ using the best linear fit of $C$ vs. $k$ in loglog space with a robust outlier correction.

\emph{Assortativity}
The degree assortativity of a network (which is often called simply `assortativity') is defined as
\begin{equation}
      a= \frac{E^{-1} \sum_{i} j_{i} k_{i} - \left[E^{-1} \sum_{i} \frac{1}{2} (j_{i} + k_{i})\right]^{2} } {E^{-1} \sum_{i} \frac{1}{2} (j_{i}^{2} + k_{i}^{2}) - \left[E^{-1} \sum_{i} \frac{1}{2} (j_{i}+k_{i})\right]^{2}}\,,
\end{equation}
where $j_{i}$ and $k_{i}$ are the degrees of the nodes at the two ends of the $i^{th}$ edge ($i \in \{1\,,\ldots\,, E\}$  \cite{Newman2006}. The assortativity measures the preference of a node to connect to other nodes of similar degree (leading to an assortative network, $r>0$) or to other nodes of very different degree (leading to a disassortative network, $r<0$). Social networks are commonly found to be assortative while networks such as the internet, World-Wide Web, protein interaction networks, food webs, and the neural network of C. elegans are disassortative.

\emph{Mean Connection Distance}
The estimated connection distance of an edge, $d_{i,j}$, is defined as the Euclidean distance between the centroids of the connected regions $i$ and $j$ in standard stereotactic space. The mean connection distance, $d$, is defined as the average connection distance over all edges in a network \citep{Bassett2008}. Thus connection distance differs from the other, topological and dimensionless network diagnostics in that it represents a spatial or topographic property of the network and has units of distance ($\mathrm{mm}$).

\emph{Rent's exponent}
Rent's exponent is a topophysical property of a network; that is, it describes how a \textit{non-physical} topology is embedded into a \textit{physical} space, which in the case of neuronal fiber tracts is the physical space of the brain \citep{Bassett2010}.  Rent's rule, which was first discovered in relation to computer chip design, defines a scaling relationship between the number of external signal connections (edges) $e$ to a block of logic and the number of connected nodes $n$ in the block \cite{Christie2000}:
\begin{equation}
         e\sim n^p\,,
\end{equation}
where $p \in [0,1]$ is the Rent exponent. Following \citep{Bassett2010}, the Rent's exponent is found by tiling the Euclidean space of the network with $N_{\mathrm{box}}=5000$ overlapping randomly sized boxes (e.g., three-dimension cubes). In each box we determine the number of nodes (\emph{n}) and the number of connections (\emph{e}) that cross the box boundaries. The gradient of a straight line fitted to $\log (n)$ versus $\log (e)$ using iteratively weighted least squares regression is an estimate of the Rent exponent $p$. To minimize boundary effects, $p$ is estimated using the subset of boxes which contains less than half the total number of nodes, $n<N/2$.

\emph{Global efficiency}
The global efficiency was defined by Latora and Marchiori \citep{Latora2001} and first applied to neuroimaging data in \citep{Achard2007}. The regional efficiency of a single node, $i$, is defined as
\begin{equation}\label{GE}
E(i) = \frac{1}{N-1} \sum_{j \in G} \frac{1}{L_{i,j}},
\end{equation}
where $i = 1,2,3,\ldots,N$ indicates the index region, $j \neq i$
denotes a region connected to $i$, and $L_{i,j}$ is the minimum path
length between regions $i$ and $j$. Regional efficiency is therefore
inversely related to minimum path length and a region with high
efficiency will have short minimum path length to all other regions
in the graph. The global efficiency of a graph is defined as the mean of $E(i)$ over all possible
regions, and is commonly denoted $E_{glob}$.

\emph{Local efficiency}
Latora and Marchiori also defined a local efficiency, which measures the efficiency of the subgraph surrounding node $i$:
\begin{equation}
E(i) = \frac{1}{N_{G_{i}}(N_{G_{i}}-1)} \sum_{j,k \in G_{i}} \frac{1}{L_{j,k}},
\end{equation}
where $G_{i}$ is the subgraph of nodes and edges connected to node $i$ and $L_{j,k}$ is the minimum path length between nodes $j$ and $k$ in the subgraph \citep{Latora2001}.

\emph{Betweenness centrality}
Geodesic node betweenness or more simply `betweenness centrality' is defined for the $i^{th}$ node in a network $\mathcal{G}$ as
\begin{equation}
            B_{i} = \sum_{j,m,i \in \mathcal{G}} \frac{\psi_{j,m}(i)}{\psi_{j,m}}\,,
            \label{eq:Bwc}
\end{equation}
where all three nodes ($j$, $m$, and $i$) must be different from each other, $\psi_{j,m}$ is the number of geodesic paths between nodes $j$ and $m$, and $\psi_{j,m}(i)$ is the number of geodesic paths between $j$ and $m$ that pass through node $i$. The betweenness centrality of an entire network $B$ is defined as the mean of $B_{i}$ over all nodes $i$ in the network.

\emph{Modularity}
Networks can be partitioned into communities or modules \cite{Porter2009,Fortunato2010} where nodes inside the same community are more densely connected to each other than they are to nodes in other communities. The modularity \cite{Girvan2002,Newman2006} of a network partition is defined as:
\begin{equation}\label{mod}
Q = \frac{1}{2m} \sum_{ij} [A_{ij} - \frac{k_{i} k_{j} }{2m}] \delta_{c_{i},c_{j}},
\end{equation}
where $k_{i}$ is the degree of node $i$, $m$ is the total number of edges in the network, $A_{ij}$ is an element of the adjacency matrix, $\delta_{ij}$ is
the Kronecker delta symbol, and $c_{i}$ is the label of the
community to which node $i$ has been assigned \citep{Leicht2008}. Here we used the Louvain locally greedy algorithm \cite{Blondel2008} to optimize the modularity quality function over the space of possible network partitions. We report the maximum value of $Q$ over this optimization procedure.

\emph{Synchronizability}
The synchronizability, $S$, of a network characterizes structural properties of a graph that hypothetically enable it to synchronize rapidly \citep{Bassett2006a}. The synchronizability is defined as
\begin{equation}
S = \frac{\lambda_{2}}{\lambda_{N}}
\end{equation}
where $\lambda_{2}$ is the second smallest eigenvalue of the Laplacian $\mathcal{L}$ of the adjacency matrix, and $\lambda_{N}$ is the largest eigenvalue of $\mathcal{L}$ \citep{Barahona2002}.

\emph{Robustness}
The robustness metric, $\rho$, indicates the network's resilience to either targeted, $\rho_{t}$, or random,
$\rho_{r}$, attack. In a targeted attack, hubs are removed one by one in order of degree, $k$, while in a random attack, nodes are removed at random independent of their degree. Each time a node was removed from the network, we re-calculated the size of the largest connected component, $s$. Robustness is then usually visualized by a
plot of the size of the largest connected component, $s$, versus the number of nodes removed, $n$ \citep{Achard2006,Lynall2010}. The robustness parameter, $\rho$, is defined as the area under this $s$ versus $n$ curve. More robust networks retain a larger connected component even when several nodes have been knocked out, as represented by a larger area under the curve or higher values of $\rho$.

\emph{Cost Efficiency}
We define the cost efficiency \citep{Achard2007,Bassett2009a,Fornito2011} at a node as the maximal difference between the regional efficiency and the network density or cost over the investigated range of cost values:
\begin{equation}\label{CE}
 CE(i) = E(i) - K.
\end{equation}
We also define the cost efficiency of the network $CE_{net}$ as the mean of $CE(i)$ over all nodes in the network.

\begin{figure*}
\begin{center}
\includegraphics[width=0.9\textwidth]{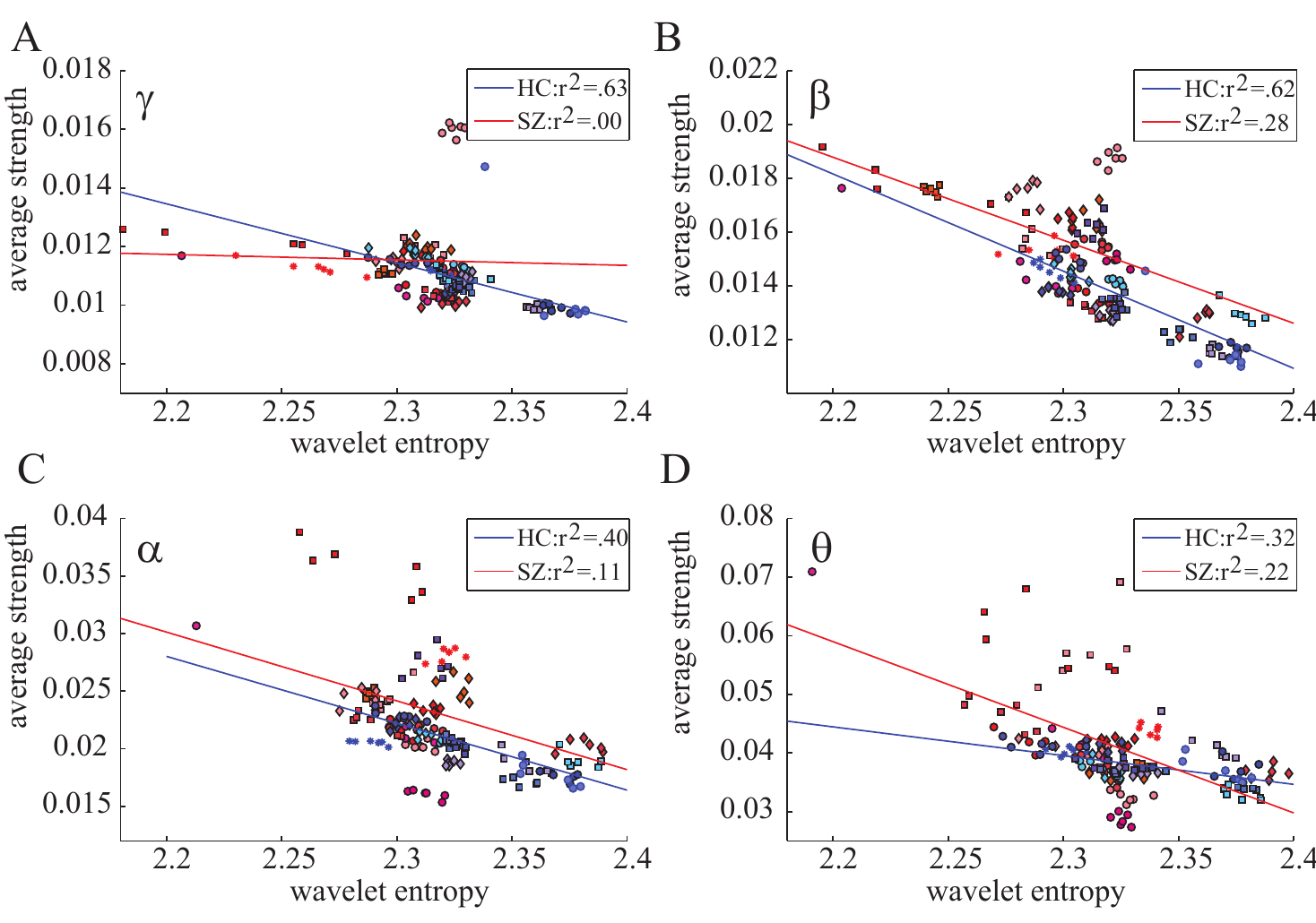}
\caption{\textbf{Entropy and Strength.}
Correlation between the strength of connectivity and complexity, as measured by wavelet entropy, in $\gamma$-, $\beta$-, $\alpha$- and $\theta$-bands. Single data points represent the mean value pairs from a block of trials, trials from different subjects are distinguished by different colors and markers. Red, orange and pink markers denote subjects with schizophrenia spectrum diagnosis; blue, turquoise and purple markers denote healthy subjects. Value pairs from single subjects exhibit a tendency to appear in clusters, which (mostly) are broken up only at low entropy and/or high connectivity. With some subjects, value pairs appear outside the main cluster in all bands, with others, only in some.
Also displayed are fitted linear functions for the two groups (red and blue lines) with $r^2$ values as indicators of goodness of fit. It should be noted that these were obtained by fitting to the mean value pairs for subjects, averaged over all 6 blocks of trials.
This was done to avoid fitting to values for which there are two sources of variance (subjects and blocks). These fits indicate a negative correlation between entropy and strength, especially for healthy subjects where $r^2$ values are much higher.
\label{fig:str_ent}}
\end{center}
\end{figure*}

\begin{figure*}
\begin{center}
\includegraphics[width=0.7\textwidth]{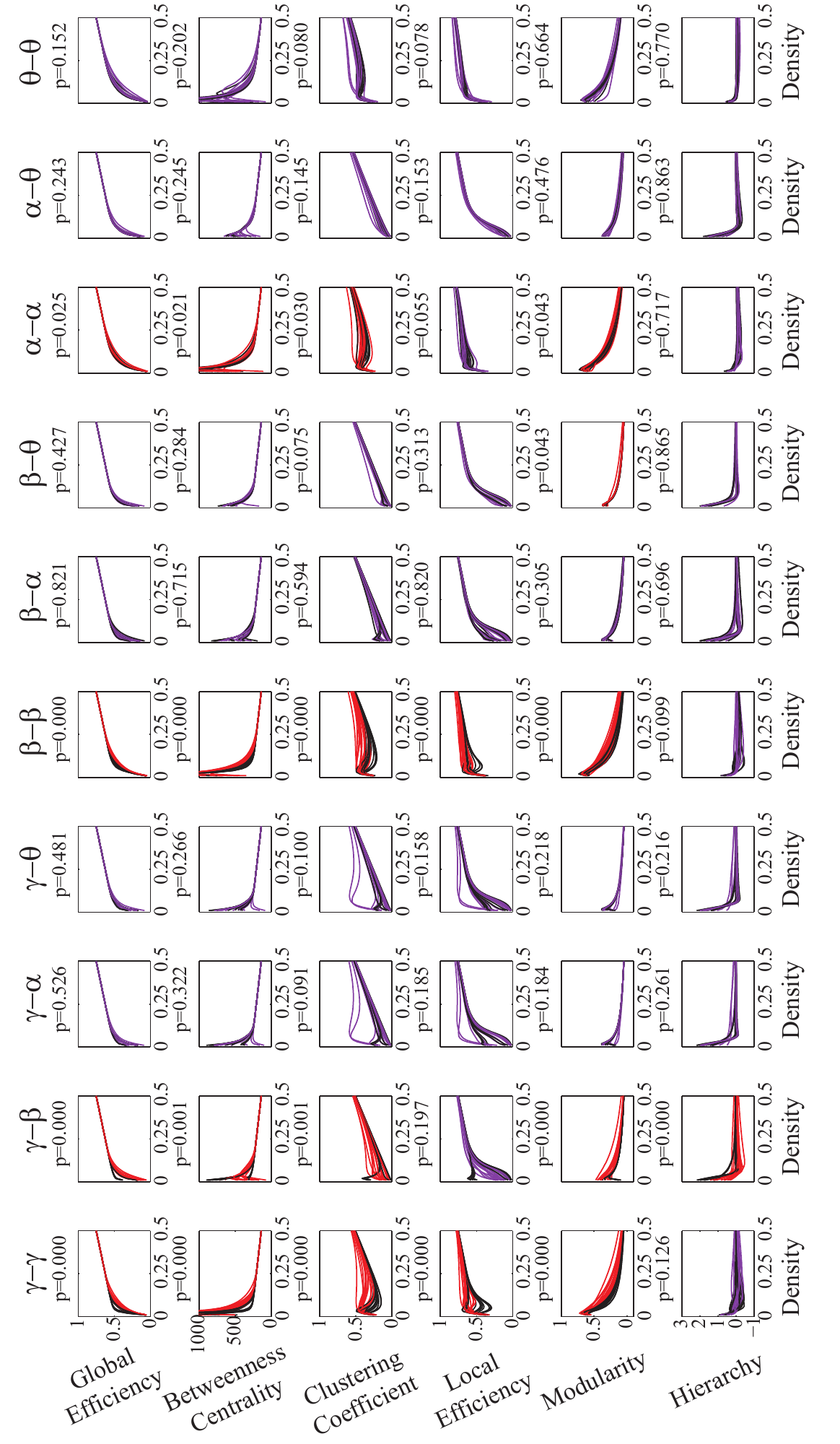}
\caption{\textbf{Network Diagnostics: Part I}
One set of six network diagnostics (global efficiency, betweenness centrality, clustering coefficient, local efficiency, modularity, and hierarchy) is plotted as a function of density in networks within and between frequency bands. Each curve represents one subject, values averaged over all 66 trials. Curves for healthy controls are black, those for SZ patients colored.
The two sets of curves were tested for statistically significant difference with Functional Data Analysis (FDA), the resulting p-values are given. Where significance ($p < 0.05$) was calculated, the color of the SZ curves was set to red, purple otherwise.
We see significant differences for most diagnostics between the groups in the $\gamma$, $\beta$ and $\alpha$ bands, as well as for the $\gamma-\beta$ cross-frequency network.
\label{fig:metrics1}}
\end{center}
\end{figure*}

\begin{figure*}
\begin{center}
\includegraphics[width=0.7\textwidth]{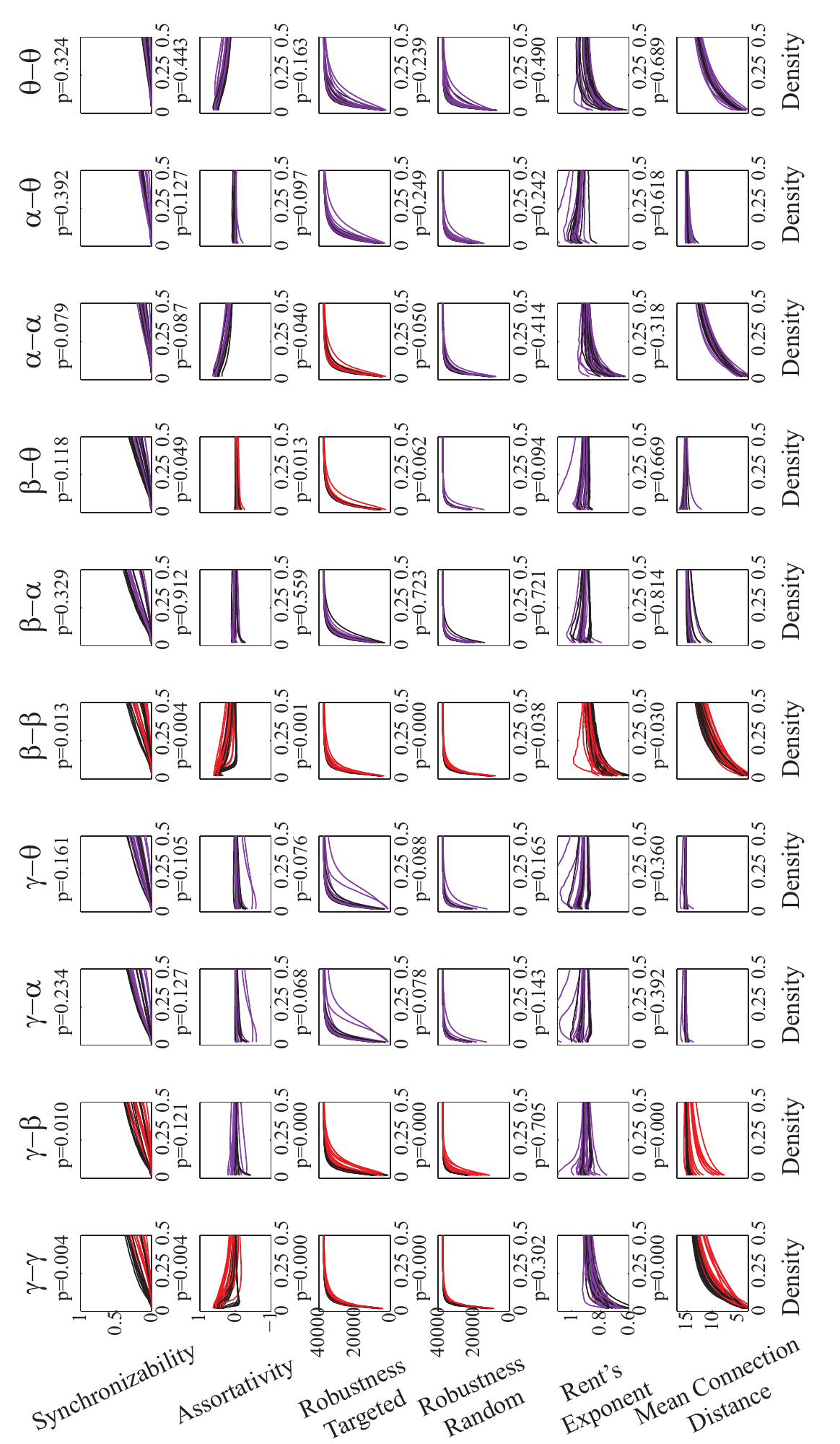}
\caption{\textbf{Network Diagnostics: Part II}
A second set of six network diagnostics (synchronizability, assortativity, robustness to targeted and random attack, Rent's exponent, and mean connection distance) is plotted as a function of density in networks within and between frequency bands. Each curve represents one subject, values averaged over all 66 trials. Curves for healthy controls are black, those for SZ patients colored.
The two sets of curves were tested for statistically significant difference with Functional Data Analysis (FDA), the resulting p-values are given. Where significance ($p < 0.05$) was calculated, the color of the SZ curves was set to red, purple otherwise. In the $\gamma$ and $\beta$ bands, as well as in the $\gamma-\beta$ cross-frequency network, we see again a majority of diagnostics showing significant ($p<0.05$) differences, but not in the $\alpha$ band.
\label{fig:metrics2}}
\end{center}
\end{figure*}

\begin{figure*}
\begin{center}
\includegraphics[width=0.9\textwidth]{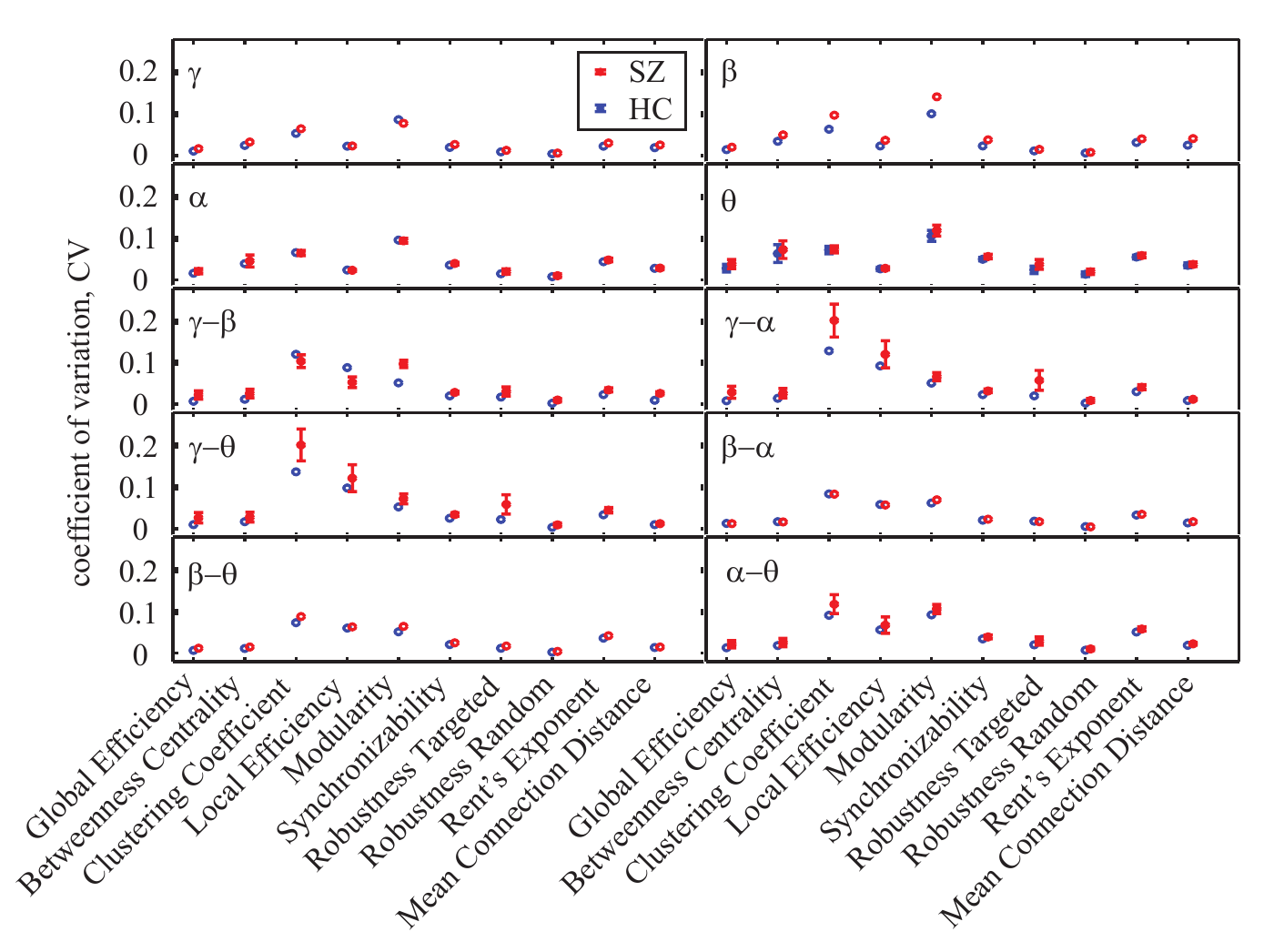}
\caption{\textbf{Variability of network diagnostics.}
Coefficient of variation for binary network diagnostics in all intra- and inter-frequency networks. Values indicate variability over trials, averaged over all healthy (blue) and schizophrenic subjects (blue) and over the entire range of cost values.
Error bars indicate the square mean of the standard errors over subjects and costs.
\label{fig:CV}}
\end{center}
\end{figure*}

\bibliographystyle{elsarticle-harv}
\bibliography{bibfile}

\end{document}